\documentclass[a4paper,reqno,10pt]{amsart} 
\usepackage{amsmath,amssymb,amsfonts}
\usepackage{latexsym,mathrsfs,cite}

\usepackage{float}
\usepackage{mathtools}
\usepackage{hyperref} 
\usepackage{placeins}
\usepackage{cleveref}
\usepackage[title,titletoc,toc]{appendix}

\usepackage{graphicx}
\usepackage{subcaption}

\numberwithin{equation}{section}
\newcounter{cont}[section]

\newtheorem{theorem}[cont]{Theorem}

\newtheorem{conjecture}[cont]{Conjecture}

\theoremstyle{definition}
\newtheorem{definition}[cont]{Definition}

\renewcommand{\Re}{\mathrm{Re}\,}
\newcommand{\cL}{\mathcal{L}}
\newcommand{\R}{\mathbb{R}}
\newcommand{\C}{\mathbb{C}}

\newcommand{\hV}{\hat{V}}
\newcommand{\hM}{\hat{M}}
\newcommand{\hu}{\hat{u}}

\newcommand{\hrho}{\hat{\rho}}
\newcommand{\hm}{\hat{m}}
\newcommand{\ess}{\sigma_\mathrm{\tiny{ess}}}

\title[Numerical spectral analysis of standing waves in QHD]{Numerical spectral analysis of standing waves in quantum hydrodynamics with viscosity}

\author[Delyan Zhelyazov]{Delyan Zhelyazov$^{\ast}$}
\thanks{$^\ast$Current address: Department of Mathematics, University of Surrey, Guildford, United Kingdom (d.zhelyazov@surrey.ac.uk)}
\address[Delyan Zhelyazov]{Departamento de Matem\'aticas y Mec\'anica\\Instituto de Investigaciones en Matem\'aticas Aplicadas y en Sistemas\\Universidad Nacional Aut\'{o}noma de M\'{e}xico\\Circuito Escolar s/n, Ciudad Universitaria, C.P. 04510\\Cd. de M\'{e}xico (Mexico)}
\theoremstyle{remark}
\newtheorem{remark}[cont]{Remark}

\begin{document}



\begin{abstract}
We study the spectrum of the linearization around standing wave profiles for two quantum hydrodynamics systems with linear and nonlinear viscosity. The essential spectrum for such profiles is stable; we investigate the point spectrum using an Evans function technique. For both systems we show numerically that there exists a real unstable eigenvalue, thus providing numerical evidence for spectral instability.
\end{abstract}

\maketitle

\section{Introduction}
In this paper, we consider two quantum hydrodynamics (QHD) systems for which we investigate the spectrum of the linearized operator about standing wave profiles. The system with a linear viscosity term reads
\begin{equation}
\label{qhd_linear_viscosity}
\begin{cases}
\rho_t + m_x = 0,\\
m_t + \left ( \frac{m^2}{\rho} + p(\rho) \right)_x = \mu m_{xx} + k^2 \rho \left ( \frac{(\sqrt \rho)_{xx}}{\sqrt{\rho}} \right )_x,
\end{cases}
\end{equation}
where $\rho \geq 0$ is the density, $m = \rho u$ is the momentum, where $u$ is the fluid velocity, $p(\rho) = \rho^{\gamma}$ with $\gamma \geq 1$ is the pressure, $t \geq 0$ and $x \in \R$. Moreover, $\mu,k > 0$ are viscosity and dispersive coefficients, respectively. The function $(\sqrt{\rho})_{xx}/\sqrt{\rho}$ is known as the Bohm potential, providing the model with a nonlinear third order dispersive term. These systems are used for instance in the theory of superfluidity \cite{Khalatnikov} or to model semiconductor devices. In the case of nonlinear viscosity (see \eqref{qhd_nonlinear_viscosity}) the term $\mu m_{xx}$ is replaced with $\mu \rho (m_x/\rho)_x$. The latter describes the interaction between the superfluid and a normal fluid, or it can be interpreted as describing the interaction of the fluid with a background. Systems \eqref{qhd_linear_viscosity} and \eqref{qhd_nonlinear_viscosity} can be viewed as mean-field limits of the de Broglie-Bohm pilot wave theory \cite{Boh52a, Boh52b, BHK87} with added viscosity terms of superfluid type.

Standing waves for \eqref{qhd_linear_viscosity} are solutions of the form
\begin{equation*}
\rho(t,x) = P(x),\mbox{ }m(t,x) = J(x),
\end{equation*}
such that
\begin{equation*}
\lim_{x \rightarrow \pm \infty} P(x) = P^{\pm} > 0,\mbox{ }\lim_{x \rightarrow \pm \infty} J(x) = J^{\pm}.
\end{equation*}

The first studies of models with dispersive terms were \cite{Sagdeev} and \cite{Gurevich1}; see also \cite{Gurevich}, \cite{Nov} and \cite{Hoefer}. Moreover, existence of weak solutions for QHD systems has been considered  in \cite{AM, AMZ, AMZ1}. The spectral stability of traveling wave profiles for the $p-$system has been discussed in \cite{Humpherys}.
Concerning viscous QHD models, existence of traveling wave profiles for the system \eqref{qhd_linear_viscosity} was shown in \cite{Zhelyazov} provided that the viscosity is sufficiently strong. The result was improved in \cite{Zhelyazov1} where global existence of dispersive shocks was proved without any restriction on the viscosity and dispersive coefficients and for arbitrary shock amplitude. A similar result about a related QHD system with nonlinear viscosity (see \eqref{qhd_nonlinear_viscosity}) was obtained in \cite{LZ}. Moreover, stability of traveling waves was established numerically in \cite{LMZ2020} and \cite{LZ2021} in the case of non-monotone shocks, and analytically in \cite{FPZ22} and \cite{FPZ23} for small-amplitude profiles. These studies were complemented in \cite{PlZ} with a proof of local well-posedness and nonlinear decay of perturbations of subsonic constant states for \eqref{qhd_linear_viscosity}.

The aim of this paper is to study the spectrum of the linearization around standing waves for \eqref{qhd_linear_viscosity} and \eqref{qhd_nonlinear_viscosity_cons}. Existence theory for such solutions was established in \cite{Zhelyazov1}. There, it was shown that the standing waves admited by the system \eqref{qhd_linear_viscosity} are pulses, i. e. $P^+ = P^-$ and $J^+ = J^-$. Here below we recall the existence result (Theorem 4.1 in \cite{Zhelyazov1}). Let us denote by $c_s(\rho) = \sqrt{\gamma \rho^{\gamma-1}}$ the speed of sound and by  $U^+ = J^+/P^+$ the velocity at the end state.
\begin{theorem}[existence of standing waves]
\label{Thm:theorem_existence}
There exists a non-constant standing wave solution of \eqref{qhd_linear_viscosity} with
\begin{equation*}
\lim_{x \rightarrow \pm \infty} [P(x), J(x)] = [P^+, J^+]
\end{equation*}
if and only if $0 < |U^+| < c_s(P^+)$.
\end{theorem}
This result shows, in particular, that there exist no standing waves with supersonic or sonic end states.

The spectrum of the linearization around standing wave profiles for systems \eqref{qhd_linear_viscosity} and \eqref{qhd_nonlinear_viscosity_cons} consists of two parts: the essential spectrum and the point spectrum.  It can be proved that the essential spectrum of such profiles is always stable (see Theorem \ref{Thm:essential_spectrum_linear} below). In the present paper, we carry out a numerical study of the point spectrum by using the Evans function method. We show that there exists an unstable real simple eigenvalue for both QHD systems with linear and nonlinear viscosity. Thereby, we provide numerical evidence for spectral instability of standing waves. This result is in contrast with the analytical and numerical results about spectral stability of traveling waves in \cite{LMZ2020, LZ2021, FPZ22, FPZ23}. A possible cause of the instability is the presence of a supersonic region along the standing wave profiles (see Section \ref{sec:discussion} and Conjecture \ref{Conj:stability_subsonic_or_sonic_profiles} below).

The paper is organized as follows. In Section \ref{sec:QHD_linear} we consider the QHD system with linear viscosity \eqref{qhd_linear_viscosity}. We introduce the equation solved by standing wave profiles and discuss their numerical computation. Then, we derive the linearization around such profiles, describe its essential spectrum, and recast the system in integrated variables. Section \ref{sec:numerics_evans_function_linear} contains the numerical result about point spectrum instability. In Section \ref{sec:QHD_nonlinear} we discuss the QHD system with nonlinear viscosity (see equation \eqref{qhd_nonlinear_viscosity_cons} below) following the steps for the linear viscosity case. Finally, we conclude the paper with a discussion of the numerical results.

\begin{remark}The linearizations in Sections \ref{sec:linearization_QHD_linear} and \ref{sec:linearization_QHD_nonlinear} can be obtained by setting $s = 0$ in the corresponding linearizations in \cite{Zhelyazov} and \cite{LZ2021} which hold also for $s = 0$. We provide them here for completeness.
\end{remark}
\section{Quantum hydrodynamics with linear viscosity}
\label{sec:QHD_linear}
\subsection{Profile equation}
In this section we shall derive the equation satisfied by a standing wave profile for system \eqref{qhd_linear_viscosity}. The Bohm potential can be rewritten in conservative form
\begin{equation*}
\rho \left ( \frac{(\sqrt{\rho})_{xx}}{\sqrt{\rho}} \right )_x = \frac{1}{2} ( \rho (\ln \rho)_{xx})_x.
\end{equation*}
Substituting the standing wave profiles $P$ and $J$ we get the following system of ODEs
\begin{align}
J' &=0, \label{ODE_cont}\\
\left ( \frac{J^2}{P} + P^{\gamma} \right )' &= \mu J'' + \frac{k^2}{2} ( P (\ln P)'')'. \label{ODE_momentum}
\end{align}
From \eqref{ODE_cont} it follows that the momentum is conserved along the profile 
\begin{equation}
\label{eq_momentum_conserved}
J(x) = A,
\end{equation}
where $A = J^+$. Since $J$ is constant we have $J'' = 0$, hence the term coming from the viscosity in equation \eqref{ODE_momentum} vanishes. Susbstituting equation \eqref{eq_momentum_conserved} in \eqref{ODE_momentum} and proceeding as in \cite[Section 2]{Zhelyazov}, we conclude that the density profile $P$ solves the ODE
\begin{equation}
\label{profile_equation}
P'' = \frac{2}{k^2} f(P) + \frac{P'^2}{P},
\end{equation}
where
\begin{equation*}
f(P) = P^{\gamma} -  B + \frac{A^2}{P},
\end{equation*}
with
\begin{equation*}
B = \left ( \frac{m^2}{\rho} + \rho^{\gamma} \right)^+.
\end{equation*}
\subsection{Numerical computation of standing wave profiles}
\label{sec:numerical_computation_profiles_linear}
We shall compute numerically standing waves whose existence is guaranteed by Theorem \ref{Thm:theorem_existence}. They correspond to homoclinic loops for the profile ODE \eqref{profile_equation}. Since the steady-state for this dynamical system to which the profile converges is a saddle, the method to compute the profile from \cite{LMZ2020} cannot be used because it relies on one of the steady-states being stable. Instead, the profile will be computed by solving a nonlinear boundary value problem.

To compute standing wave profiles we choose $P^+ > 0$ and $U^+$ that satisfy the condition of Theorem \ref{Thm:theorem_existence}. Then, we choose a sufficiently large domain size $L_1 > 0$ and integrate numerically \eqref{profile_equation} on the domain $[-L_1,L_1]$ subject to the boundary conditions
\begin{equation*}
P(-L_1) = P(L_1) = P^+,
\end{equation*}
with the boundary value solver bvp4c in Matlab. As an initial guess we use the function
\begin{equation}
\label{initial_guess}
P_0(x) = P^+ - c \exp(-x^2)
\end{equation}
for $-L_1 \leq x \leq L_1$, where $c > 0$ is a parameter to be chosen so that the solver converges. From the qualitative analysis in \cite[Section 4]{Zhelyazov1} we know that the profile has a single minimum, hence the initial guess has an appropriate shape. Using this initial guess the solver will not converge to the trivial constant solution.
\begin{figure}
\begin{center}
\includegraphics[scale=0.6]{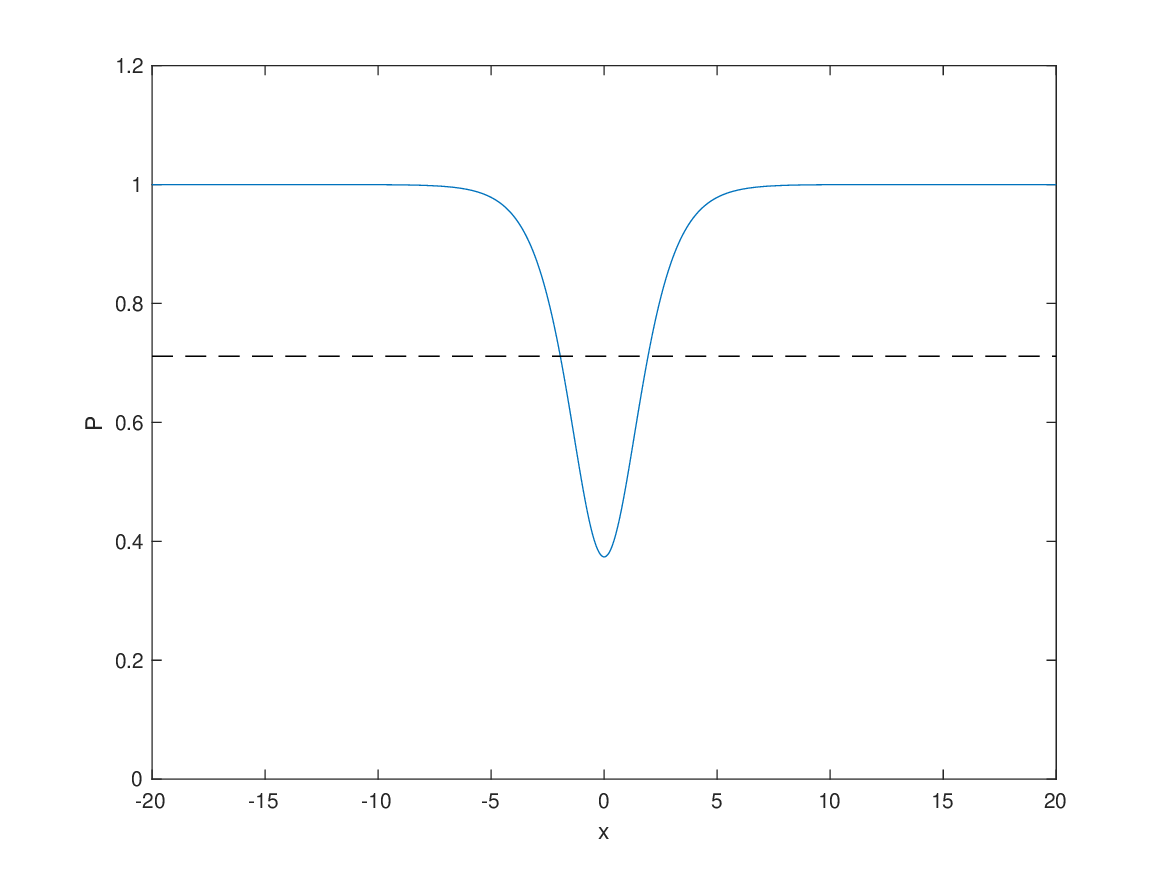}
\end{center}
\caption{Standing wave profile for parameters $P^+ = 1$, $J^+ = -0.8$, $\gamma = 3/2$, $k = \sqrt{2}$, $\mu = 0.1$. The solid line is the density profile, while the dashed line corresponds to sonic states. In the region where $P(x)$ is below the dashed line the flow is supersonic.}
\label{fig_standing_wave}
\end{figure}
Because of the fact that the momentum is conserved along the profile we have $u(x) = J^+/P(x)$. Hence, the flow is supersonic in the region where the following inequality holds
\begin{equation}
\label{cond_supersonic_flow}
P(x) < \left ( \frac{(J^+)^2}{\gamma} \right )^{\frac{1}{\gamma + 1}},
\end{equation}
where the right-hand side is the density corresponding to sonic states. For our computations we use the parameter values
\begin{equation}
\label{parameter_set}
P^+ = 1,\mbox{ }J^+ = -0.8,\mbox{ }\gamma = 3/2,\mbox{ }k = \sqrt{2},\mbox{ }\mu = 0.1.
\end{equation}
We have $|U^+| = 0.8 < 1.22 \approx c_s(P^+)$, therefore the condition of Theorem \ref{Thm:theorem_existence} is satisfied. Moreover, we set $L_1 = 20$ and $c = 0.8$. For parameters \eqref{parameter_set} the right hand side of \eqref{cond_supersonic_flow} is $0.71$ and the flow is supersonic in the interval $|x| < 1.93$ (see Figure \ref{fig_standing_wave}). This type of reduced density pulse is known as \emph{dark soliton} (see, for instance \cite{Solitons}).

\subsection{Linearization}
\label{sec:linearization_QHD_linear}
Let us denote by $(\rho, m)$ the deviation from $(P,J)$. We obtain the full linearized operator around the profile
\begin{equation}
\label{linearized_operator}
\cL
\begin{bmatrix}
\rho\\
m
\end{bmatrix}
 = \begin{bmatrix}
-m'\\
\left ( \left ( \tfrac{J^2}{P^2} - \gamma P^{\gamma - 1} \right ) \rho \right )' - \left ( \tfrac{2 J}{P}m \right )' + \mu m'' + L_V \rho 
\end{bmatrix},
\end{equation}
where
\begin{equation*}
L_V \rho = \frac{k^2}{2} \rho''' - 2 k^2 \left ( (\sqrt{P})' \left ( \frac{\rho}{\sqrt{P}} \right )' \right ) '.
\end{equation*}
The associated eigenvalue problem reads
\begin{equation}
\label{eigenvalue_equation}
\cL \begin{bmatrix}
\rho\\
m
\end{bmatrix} = 
\lambda \begin{bmatrix}
\rho\\
m
\end{bmatrix}.
\end{equation}
\subsection{Essential Spectrum}
Since the profile is a pulse, the essential spectrum of \eqref{linearized_operator} is obtained by considering the asymptotic operator at the end state
\begin{equation*}
\cL_{\infty} \begin{bmatrix}
\rho\\
m
\end{bmatrix} = 
\begin{bmatrix}
- m'\\
\alpha^+ \rho' + \beta^+ m' + \mu m'' + \frac{k^2}{2}\rho'''
\end{bmatrix},
\end{equation*}
where
\begin{equation*}
\alpha^+ = \frac{(J^+)^2}{(P^+)^2} - \gamma (P^+)^{\gamma - 1},\mbox{ }\beta^+ = - \frac{2 J^+}{P^+}.
\end{equation*}
The associated eigenvalue problem is
\begin{equation*}
\cL_{\infty} \begin{bmatrix}
\rho\\
m
\end{bmatrix}
=
\lambda \begin{bmatrix}
\rho\\
m
\end{bmatrix}.
\end{equation*}
Let us rewrite it as a first order system
\begin{equation*}
V' = M^+ V,
\end{equation*}
where $V = [\rho, m, u_1, u_2]^{\top}$, with $\rho' = u_1$, $u_1' = u_2$, and the limit matrix $M^+$ is given by
\begin{equation}
\label{mat_Mp}
M^+ = \begin{bmatrix}
0 & 0 & 1 & 0\\
-\lambda & 0 & 0 & 0\\
0 & 0 & 0 & 1\\
\tfrac{2 \beta^+ \lambda}{k^2} & \tfrac{2 \lambda}{k^2} & \tfrac{2}{k^2}(\mu \lambda - \alpha^+) & 0
\end{bmatrix}.
\end{equation}
The characteristic equation of $M^+$ is $\det(\nu Id - M^+) = 0$. It reads
\begin{equation}
\label{characteristic_equation}
\nu^4 + \frac{2}{k^2}(\alpha^+ - \lambda \mu)\nu^2 - 2 \frac{\beta^+}{k^2}\lambda \nu + \frac{2 \lambda^2}{k^2} = 0.
\end{equation}
Setting $\nu = i \xi$, $\xi \in \R$ in \eqref{characteristic_equation} and dividing by $2/k^2$ we obtain the dispersion relation
\begin{equation}
\label{dispersion_relation}
\lambda^2 + \xi (\mu \xi - i \beta^+)\lambda - \xi^2 \left ( \alpha^+ - \frac{k^2 \xi^2}{2} \right ) = 0.
\end{equation}
Since the profile $(P,J)$ is a pulse, the essential spectrum is given by the union of the two curves $\Sigma_j = \{ \lambda = \lambda_j(\xi) \in \C: \xi \in \R \}$, $j = 1,2$, where $\lambda_j(\xi)$ are determined by the roots of \eqref{dispersion_relation}. Let us now define stability of essential spectrum.
\begin{definition}
The essential spectrum of $\cL$ is \emph{stable}, if it is contained in the closed left half-plane, that is
\begin{equation*}
\ess(\cL) \subseteq \{ \lambda \in \C: \Re \lambda \leq 0 \}.
\end{equation*}
The essential spectum of $\cL$ is \emph{unstable}, if it intersects the unstable half-plane:
\begin{equation*}
\ess(\cL) \cap \{ \lambda \in \C: \Re \lambda > 0 \} \neq \varnothing.
\end{equation*}
\end{definition} Lemma 3 in \cite{Zhelyazov} and Lemma 3.2 in \cite{PlZ} imply the following description of the stability of the essential spectrum.
\begin{theorem}[Stability of essential spectrum]
\label{Thm:essential_spectrum_linear}
\begin{itemize}
\item[]
\item[\rm{(i)}] If $|U^+| \leq c_s(P^+)$, then the essential spectrum of $\cL$ is stable. Moreover, it holds that $\Re \lambda_{1,2} < 0$, provided $\xi \neq 0$.
\item[\rm{(ii)}] If $|U^+| > c_s(P^+)$, then the essential spectum of $\cL$ is unstable.
\end{itemize}
\end{theorem}
That is, the stability of the essential spectrum is related to the end state $(P^+,J^+)$ being subsonic or sonic. Moreover, the existence condition in Theorem \ref{Thm:theorem_existence} is also related to stability of the essential spectrum. Indeeed, a standing wave profile exists if and only if the end state is subsonic with nonzero velocity. Hence, the essential spectrum of standing waves is always stable. Furthermore, Lemma 3 in \cite{Zhelyazov} implies that we have consistent splitting, that is the limit matrix $M^+$ has 2 eigenvalues with positive real parts and 2 eigenvalues with negative real parts provided $\lambda$ is to the right of the curves $\Sigma_j$.
\subsection{System in integrated variables}
Following \cite{Zhelyazov} (see Section 4.2.1), we shall recast the eigenvalue problem \eqref{eigenvalue_equation} in terms of integrated variables
\begin{equation*}
\hrho(x) = \int_{-\infty}^x \rho(y) dy,\mbox{ }\hm(x) = \int_{-\infty}^x m(y) dy.
\end{equation*}
This transformation removes the zero eigenvalue without further modifications of the spectrum. Expressing $\rho$ and $m$ in terms of $\hrho$ and $\hm$ and integrating from $-\infty$ to $x$ we get
\begin{align}
\lambda \hrho &= - \hm', \label{sys_integ_vars_1}\\
\lambda \hm &= f_1 \hrho' + f_2 \hm' + \frac{k^2}{2}\hrho''' -2 k^2 (\sqrt{P})'\left ( \frac{\hrho'}{\sqrt{P}} \right )', \label{sys_integ_vars_2}
\end{align}
where
\begin{equation*}
f_1(x) = \frac{J(x)^2}{P(x)^2} - \gamma P(x)^{\gamma - 1},\mbox{ }f_2(x) = -2 \frac{J(x)}{P(x)}.
\end{equation*}
Let us rewrite the system \eqref{sys_integ_vars_1}-\eqref{sys_integ_vars_2} as $\hV' = \hM(x,\lambda)\hV$, where $\hV = [\hrho, \hm, \hu_1, \hu_2]^{\top}$, $\hrho' = \hu_1$, $\hu_1' = \hu_2$, and
\begin{equation}
\label{def_mat_integ_vars}
\hM(x,\lambda) = \begin{bmatrix}
0 & 0 & 1 & 0\\
-\lambda & 0 & 0 & 0\\
0 & 0 & 0 & 1\\
\tfrac{2 \lambda f_2}{k^2} & \tfrac{2 \lambda}{k^2} & \tfrac{2 \lambda \mu}{k^2} - \tfrac{2 f_1}{k^2} - \tfrac{P'^2}{P^2} & \tfrac{2 P'}{P}
\end{bmatrix}.
\end{equation}
The limiting matrix of $\hM(x,\lambda)$ as $x \rightarrow \pm \infty$ is \eqref{mat_Mp}.
\subsection{Numerical evaluation of the Evans function}
\label{sec:numerics_evans_function_linear}
\begin{figure}
\begin{center}
\includegraphics[scale=0.6]{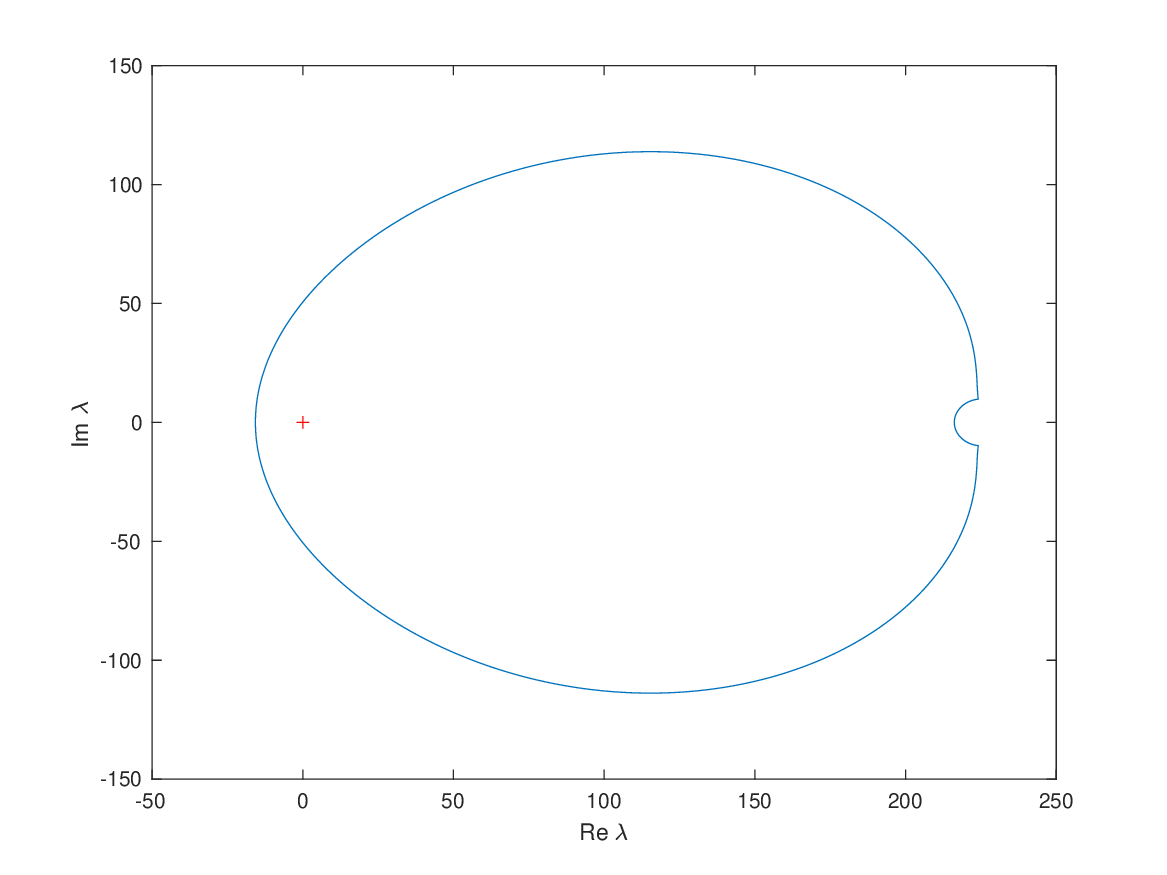}
\end{center}
\caption{The image of a semi-circular contour with radius 20 through the Evans function $E(\lambda)$ for the QHD system with linear viscosity. The origin is marked in red.}
\label{fig_evans_function}
\end{figure}

To define the Evans function, let us consider the equation $Y' = \hM(x,\lambda)Y$ with $\hM(x,\lambda)$ defined in \eqref{def_mat_integ_vars}. Since the profile $(P,J)$ is a pulse, the limits of $\hM(x,\lambda)$ at $\pm \infty$ coincide and are equal to the matrix $M^+$ given by \eqref{mat_Mp}. We assume that $M^+$ is hyperbolic. This is always true for $\lambda$ to the right of the curves $\Sigma_j$. Denote by $\nu^-_1,\nu^-_2$ and by $\nu^+_1,\nu^+_2$ the eigenvalues of $M^+$ with positive and negative real parts, respectively, and indicate with $v_i^{\pm}$ the corresponding (normalized) eigenvectors. Let $Y^-_i$ be a solution of $Y'=\hM(x,\lambda)Y$, satisfying $\exp(\nu^-_i x)Y^-(x)$ tends to $v^-_i$ as $x \rightarrow - \infty$ and $\exp(\nu^+_i x)Y^+(x)$ tends to $v^+_i$ as  $x \rightarrow +\infty$. Then, the Evans function can be defined by
\begin{equation*}
E(\lambda) = \det(Y^-_1(0), Y^-_2(0),Y^+_1(0),Y^+_2(0)).
\end{equation*}
Consequently, $\lambda$ is in the point spectrum of $\cL$ if and only if $E(\lambda) = 0$.\\
\\
To numerically compute the Evans function, we use the compound matrix method (see \cite{Humpherys} and \cite{LMZ2020}, \cite{LZ2021}).
This method  is used in order to get a stable numerical procedure, in spite of the fact that the system $Y'=\hM(x,\lambda)Y$ is numerically stiff.
Specifically, the compound matrix $B(x,\lambda)$ is given by:\
\begin{equation*}
B=\begin{bmatrix}
    m_{11}+m_{22} & m_{23} & m_{24} & -m_{13} & -m_{14} & 0 \\
    m_{32} & m_{11}+m_{33} & m_{34} & m_{12} & 0 & -m_{14} \\
    m_{42} & m_{43} & m_{11}+m_{44} & 0 & m_{12} & m_{13} \\
    -m_{31} & m_{21} & 0 & m_{22}+m_{33} & m_{34} & -m_{24} \\
    -m_{41} & 0 & m_{21} & m_{43} & m_{22}+m_{44} & m_{23} \\
     0 & -m_{41} & m_{31} & -m_{42} & m_{32} & m_{33}+m_{44}
  \end{bmatrix},
\end{equation*}
where $m_{jk}$ are the entries of $\hM(x,\lambda)$ defined in \eqref{def_mat_integ_vars}. We integrate the equation $\phi'=(B(x,\lambda)-\mu^-)\phi$ numerically  on a sufficiently large interval $[-L_1,0]$, where $\mu^-$ is the unstable eigenvalue of $B$ at $-\infty$ with maximal (positive) real part. Denote the profile $(P(x), J(x))$ by $\zeta(x)$. Given a numerical approximation $(\zeta_k)_{k=1}^N$ of $\zeta(x)$ at points $(x_k)_{k=1}^N$ with $-L_1 = x_1 < x_2 < ... < x_N = L_1$, let $\tilde{\zeta}(x)$ be the piecewise linear interpolant of $(x_1,\zeta_1),...,(x_N,\zeta_N)$. We obtain the matrix $B(x,\lambda)$ using $\tilde{\zeta}(x)$.
Similarly we integrate the equation $\phi'=(B(x,\lambda)-\mu^+)\phi$ on $[0,L_1]$ backwards, where this time  $\mu^+$ is the stable eigenvalue of $B$ at $+\infty$ with minimal (negative) real part. Then, the coefficients $\mu^{\pm}$ compensate for the growth/decay at infinity. Finally, the Evans function can be constructed by means of linear combination of the components of the two solutions $\phi^\pm = (\phi^\pm_1, \dots, \phi^\pm_6)$ as follows:
\begin{equation*}
E(\lambda)=\phi^-_1\phi^+_6-\phi^-_2\phi^+_5+\phi^-_3\phi^+_4+\phi^-_4\phi^+_3-\phi^-_5\phi^+_2+\phi^-_6\phi^+_1\Big |_{x=0}.
\end{equation*}
To compute the initial conditions we integrate the reduced Kato ODE
\begin{equation}
\label{reduced_kato_ODE}
\frac{d r_{\pm}}{d \lambda} = \frac{d\mathcal{P}_{\pm}}{d\lambda}r_{\pm},
\end{equation}
where $\mathcal{P}_{\pm}$ denotes the spectral projection of $\hat{B}_{\pm} = \lim_{x \rightarrow + \infty} \hat{B}(x,\lambda)$ corresponding to $\mu^{\pm}$. This choice of initial condition provides an analytic Evans function $E(\lambda)$. To numerically integrate \eqref{reduced_kato_ODE} we use the algorithm from \cite{Zumbrun}, that is $|r_{\pm}^1| = 1$ eigenvector as before (referring to maximal/minimal decay/growth rate of $\hat{B}^{\pm}$) and for $k > 0$,
\begin{equation*}
r_{\pm}^{k+1} = \mathcal{P}_{\pm}^k r_{\pm}^k.
\end{equation*}
Then, provided $E(\lambda)$ does not vanish on a closed contour $\Gamma$ which does not intersect the essential spectrum of $\mathcal{L}$, we can use the winding number
\begin{equation*}
\frac{1}{2 \pi i}\int_{\Gamma}\frac{E'(z)}{E(z)}dz
\end{equation*}
to count the number of zeros (taking into account multiplicity) inside the contour (see \cite{Sandstede}).\\
\\
For our calculations we use the set of parameters \eqref{parameter_set}. Recall that the associated profile is depicted in Figure \ref{fig_standing_wave}. We compute the Evans function on a semi-circular contour with radius 20, center at $\lambda = 0$ and vertical segment on the imaginary axis. Furthermore, we do not evaluate the Evans function at 0, but evaluate it up to $\pm i 10^{-6}$. Along the contour we integrate the Kato ODE using $5 \cdot 10^4$ points. Then, we compute $E(\lambda)$ with the solver ode45 in Matlab with relative tolerance $10^{-6}$ and we set $L_1 = 40$. Finally, we apply the symmetry of
\begin{equation}
\label{evans_function_symmetry}
E(\overline{\lambda}) = \overline{E(\lambda)}.
\end{equation}
The Evans function $E(\lambda)$ is plotted in Figure \ref{fig_evans_function}. Contrary to the case of traveling waves studied in \cite{LMZ2020} and \cite{LZ2021}, the winding number of the Evans function is (approximately) 1. This is a numerical evidence for the presence of an unstable real simple eigenvalue in the interval $(0,20)$. Indeed, if $\nu$ would be a root of $E(\lambda) = 0$ with nonzero imaginary part, then by \eqref{evans_function_symmetry} $\overline{\nu}$ would also be a root, and hence the winding number would be greater than 1. Therefore, there exists a real root with multiplicity 1.

\begin{figure}
\begin{center}
\includegraphics[scale=0.6]{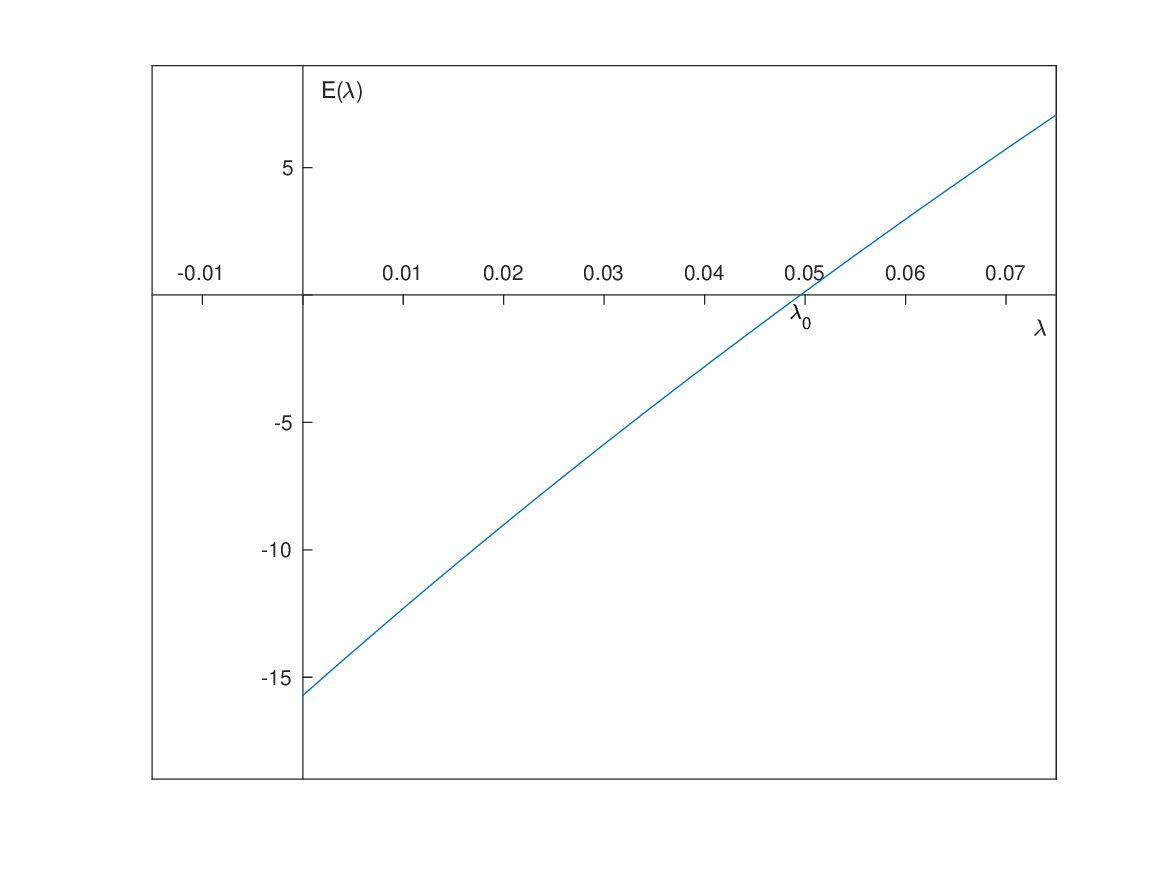}
\end{center}
\caption{The Evans function $E(\lambda)$ on the real line for the QHD system with linear viscosity. It intersects the horizontal axis at $\lambda_0 \approx 0.0496$.}
\label{fig_evans_function_real}
\end{figure}

Let us compute $E(\lambda)$ on the real line. Equation \eqref{evans_function_symmetry} implies that $E(\lambda) \in \R$ for $\lambda \in \R$. We initialize the Kato ODE at $\lambda = 20$ and integrate it along the real line until $\lambda = 10^{-6}$ using $2 \cdot 10^4$ points. The result is plotted in Figure \ref{fig_evans_function_real}. The Evans function has a real root $\lambda_0 \approx 0.0496$.

The instability possibly is related to the presence of a supersonic region along the profile as will be discussed in Section \ref{sec:discussion}.

\section{Quantum Hydrodynamics with Nonlinear Viscosity}
\label{sec:QHD_nonlinear}
In this section we consider the following quantum hydrodynamics system with nolinear viscosity
\begin{equation}
\label{qhd_nonlinear_viscosity}
\begin{cases}
\rho_t + m_x = 0,\\
m_t + \left ( \frac{m^2}{\rho} + p(\rho) \right)_x = \mu \rho \left ( \frac{m_x}{\rho} \right)_x + k^2 \rho \left ( \frac{(\sqrt \rho)_{xx}}{\sqrt{\rho}} \right )_x.
\end{cases}
\end{equation}
Let us rewrite system \eqref{qhd_nonlinear_viscosity} in terms of $(\rho, u)$ variables as
\begin{equation}
\label{qhd_nonlinear_viscosity_cons}
\begin{cases}
\rho_t + (\rho u)_x = 0,\\
u_t + \frac{(u^2)_x}{2} + (h(\rho))_x = \mu \left ( \frac{(\rho u)_x}{\rho} \right )_x + k^2 \left ( \frac{(\sqrt{\rho}_{xx})}{\sqrt{\rho}} \right )_x,
\end{cases}
\end{equation}
where the enthalpy $h(\rho)$ is
\begin{equation*}
h(\rho) = \begin{cases}
\ln \rho, &\gamma = 1,\\
\frac{\gamma}{\gamma - 1}\rho^{\gamma - 1}, &\gamma > 1,
\end{cases}
\end{equation*}
see \cite{Gasser}. We are searching for standing wave profiles of the form
\begin{equation}
\label{standing_wave_profiles_nonlinear}
\rho(t,x) = R(x),\mbox{ }u(t,x) = U(x),
\end{equation}
such that
\begin{equation*}
\lim_{x \rightarrow \pm \infty}R(x) = R^+,\mbox{ }\lim_{x \rightarrow \pm \infty}U(x) = U^+.
\end{equation*}
Substituting \eqref{standing_wave_profiles_nonlinear} into \eqref{qhd_nonlinear_viscosity_cons} we obtain
\begin{align}
(R U)' &= 0, \label{profile_eq_nonlinear_1}\\
\frac{1}{2}(U^2)' + h(R)' &= \mu \left ( \frac{(R U)'}{R} \right)' + k^2 \left ( \frac{(\sqrt{R})''}{\sqrt{R}} \right )'. \label{profile_eq_nonlinear_2}
\end{align}
Integrating equation \eqref{profile_eq_nonlinear_1} up to $\pm \infty$ we get
\begin{equation}
\label{expr_U}
U = \frac{A}{R},
\end{equation}
where $A = R^+ U^+$. Substituting \eqref{expr_U} into \eqref{profile_eq_nonlinear_2} and integrating we obtain
\begin{equation}
\label{eq_second_order_nonlinear_viscosity}
R'' = \frac{2}{k^2}f(R) + \frac{(R')^2}{2 R},
\end{equation}
where
\begin{equation*}
f(R) = R h(R) + \frac{A^2}{2 R} - R B,
\end{equation*}
with
\begin{equation*}
B = \frac{1}{2}(U^+)^2 + h(R^+).
\end{equation*}
Theorem 6.1 in \cite{Zhelyazov1} provides a characterization for the existence of homoclinic loops to \eqref{eq_second_order_nonlinear_viscosity}. These homoclinic loops correspond to standing wave profiles for \eqref{qhd_nonlinear_viscosity_cons}.

\subsection{Numerical computation of profiles}
\begin{figure}
\begin{center}
\includegraphics[scale=0.6]{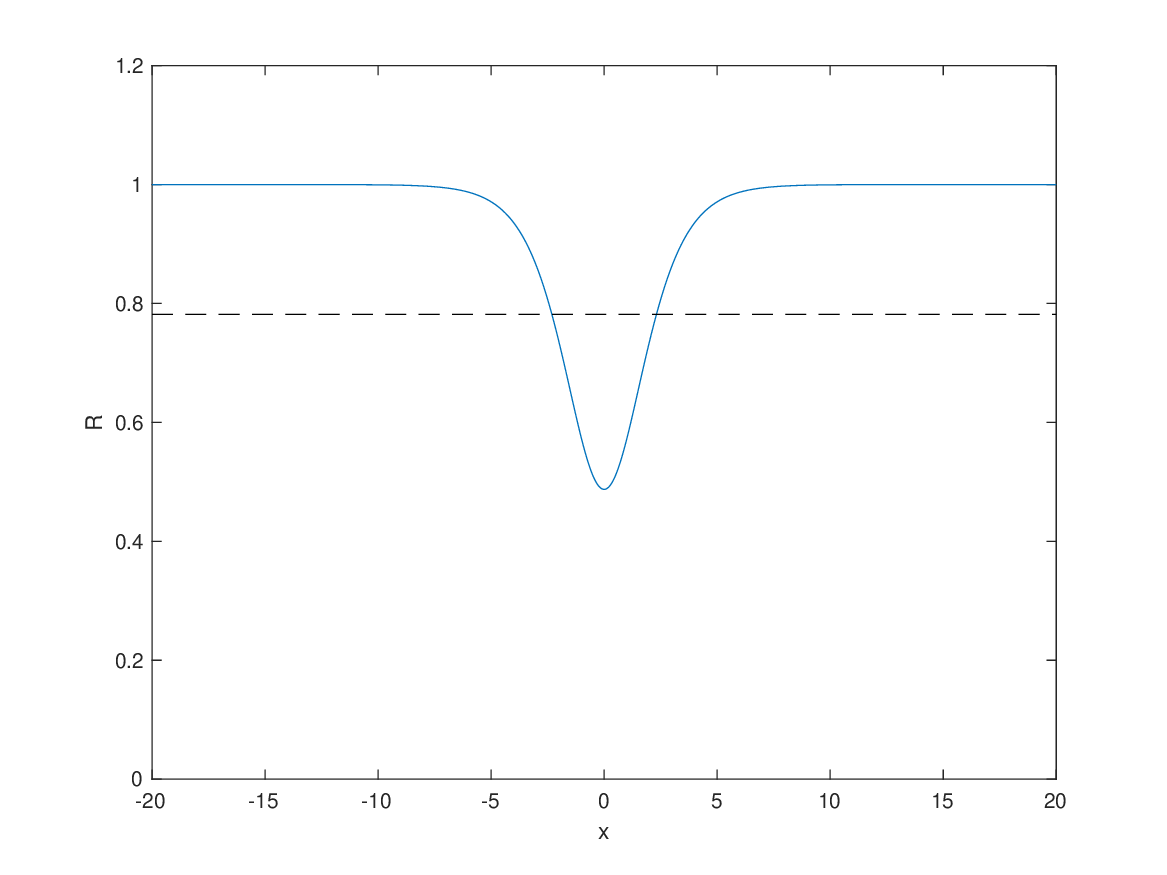}
\end{center}
\caption{Standing wave profile for parameters $R^+ = 1$, $U^+ = 0.9$, $\gamma = 3/2$, $k = \sqrt{2}$, $\mu = 0.08$. The solid line corresponds to $R(x)$. In the region where $R(x)$ is below the dashed line the flow is supersonic.}
\label{fig_standing_wave2}
\end{figure}

We use the method from Section \ref{sec:numerical_computation_profiles_linear}, that is we compute the profile numerically using a boundary value solver. As an initial guess we use a function of the form \eqref{initial_guess} with $c = 0.5$,
\begin{equation*}
R_0(x) = R^+ - 0.5 \exp(-x^2).
\end{equation*}
We compute the profile for parameters
\begin{equation}
\label{parameter_set_nonlinear_viscosity}
R^+ = 1,\mbox{ }U^+ = 0.9,\mbox{ }\gamma = 3/2,\mbox{ }k = \sqrt{2},\mbox{ }\mu = 0.08.
\end{equation}
Since $|U^+| = 0.9 < 1.22 \approx c_s(R+)$, the condition for existence of profile of Theorem 6.1 in \cite{Zhelyazov1} is verified;
see Figure \ref{fig_standing_wave2}. The flow along the profile is supersonic in the region where the following inequality holds:
\begin{equation}
\label{supersonicity_condition_nonlinear}
R(x) < \left ( \frac{(R^+ U^+)^2}{\gamma} \right )^{\frac{1}{\gamma + 1}}.
\end{equation}
For the parameter values \eqref{parameter_set_nonlinear_viscosity} the right hand side of \eqref{supersonicity_condition_nonlinear} is 0.78 and the flow is supersonic for $|x| < 2.29$.
\subsection{Linearization}
\label{sec:linearization_QHD_nonlinear}
Here we perform a linearization of \eqref{qhd_nonlinear_viscosity_cons} around a standing wave profile. Denoting by $(\rho, u)$ the deviation from the profile $(R, U)$, we obtain the following full linearized operator:
\begin{equation*}
\cL \begin{bmatrix}
\rho \\ u
\end{bmatrix}
:= \begin{bmatrix}
-(R u + U \rho)' \\
-(U u)' - \left ( \dfrac{d h}{d R}(R) \rho \right )'\\
+ \mu \left ( (R^{-1} (R u + U \rho)')' - (R^{-2}(R U)' \rho)'  \right ) + k^2 \cL_Q \rho
\end{bmatrix},
\end{equation*}
where
\begin{equation*}
\cL_Q \rho = \frac{1}{2} ( R^{-1/2} (R^{-1/2} \rho)'' )' - \frac{1}{2} (R^{-3/2} (R^{1/2})'' \rho)'.
\end{equation*}
The associated eigenvalue problem reads
\begin{equation}
\label{eigenvalue_equation_nonlinear_viscosity}
\lambda \begin{bmatrix}
\rho \\ u
\end{bmatrix} = 
\cL
\begin{bmatrix}
\rho \\ u
\end{bmatrix}.
\end{equation}

The asymptotic operator at $\infty$ is given by
\begin{equation*}
\cL_{\infty} \begin{bmatrix}
\rho \\ u
\end{bmatrix} :=
\begin{bmatrix}
- U^+ \rho' - R^+ u'\\
-U^+ u' - \dfrac{dh}{dR}(R^+) \rho' + \mu \left ( u'' + \dfrac{U^+}{R^+} \rho'' \right ) + \dfrac{k^2}{2} \dfrac{\rho'''}{R}
\end{bmatrix}.
\end{equation*}
The eigenvalue problem associated to $\cL_{\infty}$ is
\begin{equation*}
\lambda \begin{bmatrix}
\rho \\ u
\end{bmatrix} =
\cL_{\infty} \begin{bmatrix}
\rho \\ u
\end{bmatrix}.
\end{equation*}
Let us rewrite it as a first-order system
\begin{equation*}
V' = M^+ V,
\end{equation*}
where
\begin{equation*}
M^+ := \begin{bmatrix}
0 & 0 & 1 & 0\\
-\dfrac{\lambda}{R^+} & 0 & - \dfrac{U^+}{R^+} & 0 \\
0 & 0 & 0 & 1\\
-\dfrac{2 U^+ \lambda}{k^2} & \dfrac{2 R^+ \lambda}{k^2} & \dfrac{2}{k^2} \left ( R^+ \dfrac{dh}{dR}(R^+) - (U^+)^2 + \mu \lambda \right ) & 0
\end{bmatrix}.
\end{equation*}
\subsection{Essential Spectrum}
The characteristic equation $\det (\nu Id - M^+) = 0$ is
\begin{equation}
\label{char_eq_nonlinear}
\nu^4 + \frac{2}{k^2} \left ((U^+)^2 - R^+\frac{dh}{dR}(R^+) - \lambda \mu \right ) \nu^2 + \frac{4 U^+}{k^2} \lambda \nu + \frac{2 \lambda^2}{k^2} = 0.
\end{equation}
Setting $\nu = i \xi$, $\xi \in \R$, in \eqref{char_eq_nonlinear} and dividing by $2/k^2$, we obtain the dispersion relation:
\begin{equation*}
\lambda^2 + (\mu \xi^2 + 2 U^+ \xi i) \lambda + \left ( R^+ \frac{dh}{dR}(R^+) - (U^+)^2 \right ) \xi^2 + \frac{k^2}{2} \xi^4 = 0.
\end{equation*}
The stability condition for the essential spectrum is identical to Theorem \ref{Thm:essential_spectrum_linear}. Therefore, standing wave profiles for the system \eqref{qhd_nonlinear_viscosity_cons} always have stable essential spectrum.
\subsection{System in integrated variables}
Let us now consider the integrated variables
\begin{equation*}
\hrho(x) = \int_{-\infty}^x \rho(y) dy,\mbox{ }\hu(x) = \int_{-\infty}^x u(y) dy.
\end{equation*}
Expressing $\rho$ and $u$ in terms of $\hrho$ and $\hu$ in \eqref{eigenvalue_equation_nonlinear_viscosity} and integrating from  $-\infty$ to $x$ we obtain the system in integrated variables:
\begin{align}
\lambda \hrho &= - U \hrho' - R \hu', \label{sys_integ_vars_nonlinear_visc_1}\\
\lambda \hu &= f(x) \rho' - U u' + \mu ( R^{-1} ( R \hu' + U \hrho')' - R^{-2}(R U)' \hrho') \nonumber\\
&+ \frac{k^2}{2}(R^{-\frac{1}{2}} (R^{-\frac{1}{2}} \hrho')'' - R^{-\frac{3}{2}}(R^\frac{1}{2})'' \hrho'), \label{sys_integ_vars_nonlinear_visc_2}
\end{align}
with
\begin{equation*}
f(x) = - \frac{dh}{dR}(R(x)).
\end{equation*}
Let us rewrite \eqref{sys_integ_vars_nonlinear_visc_1}-\eqref{sys_integ_vars_nonlinear_visc_2} as a first-order system
\begin{equation*}
\hV' = \hM(x,\lambda)\hV,
\end{equation*}
where $\hV = [\hrho, \hu, \hu_1, \hu_2]^{\top}$, with $\hrho' = \hu_1$, and $\hu_1' = \hu_2$, and
\begin{equation}
\label{mat_integ_vars_nonlinear}
\hM(x, \lambda) := \begin{bmatrix}
0 & 0 & 1 & 0\\
-\tfrac{\lambda}{R} & 0 & -\tfrac{U}{R} & 0\\
0 & 0 & 0 & 1\\
\hm_{4,1} & \hm_{4,2} & \hm_{4,3} & \hm_{4,4}
\end{bmatrix},
\end{equation}
with
\begin{align*}
\hm_{4,1} &= - \frac{2 \lambda U}{k^2},\\
\hm_{4,2} &= \frac{2 \lambda R}{k^2},\\
\hm_{4,3} &= \frac{2}{k^2} \left ( - R f- U^2 + \mu \left ( \frac{(R U)'}{R} + \lambda \right) \right) + \frac{R''}{R} - \frac{(R')^2}{R^2},\\
\hm_{4,4} &= \frac{R'}{R}.
\end{align*}
\subsection{Numerical evaluation of the Evans function}
\label{sec:numerical_evans_function_nonlinear}
In this section we compute the Evans function for the QHD system with nonlinear viscosity \eqref{qhd_nonlinear_viscosity_cons}. We use the method from Section \ref{sec:numerics_evans_function_linear}. We integrate the ODE $\phi'=(B(x,\lambda)-\mu^-)\phi$ on $[-L_1,0]$ and $\phi'=(B(x,\lambda)-\mu^+)\phi$ on $[0,L_1]$ backwards, where $L_1 = 40$ as before, however here $B(x,\lambda)$ is the compound matrix of $\hM(x,\lambda)$ defined in \eqref{mat_integ_vars_nonlinear} and $\mu^{\pm}$ is the stable (unstable) eigenvalue of $B$ at $+\infty$ with minimal (maximal) real part. 

\begin{figure}
\begin{center}
\includegraphics[scale=0.6]{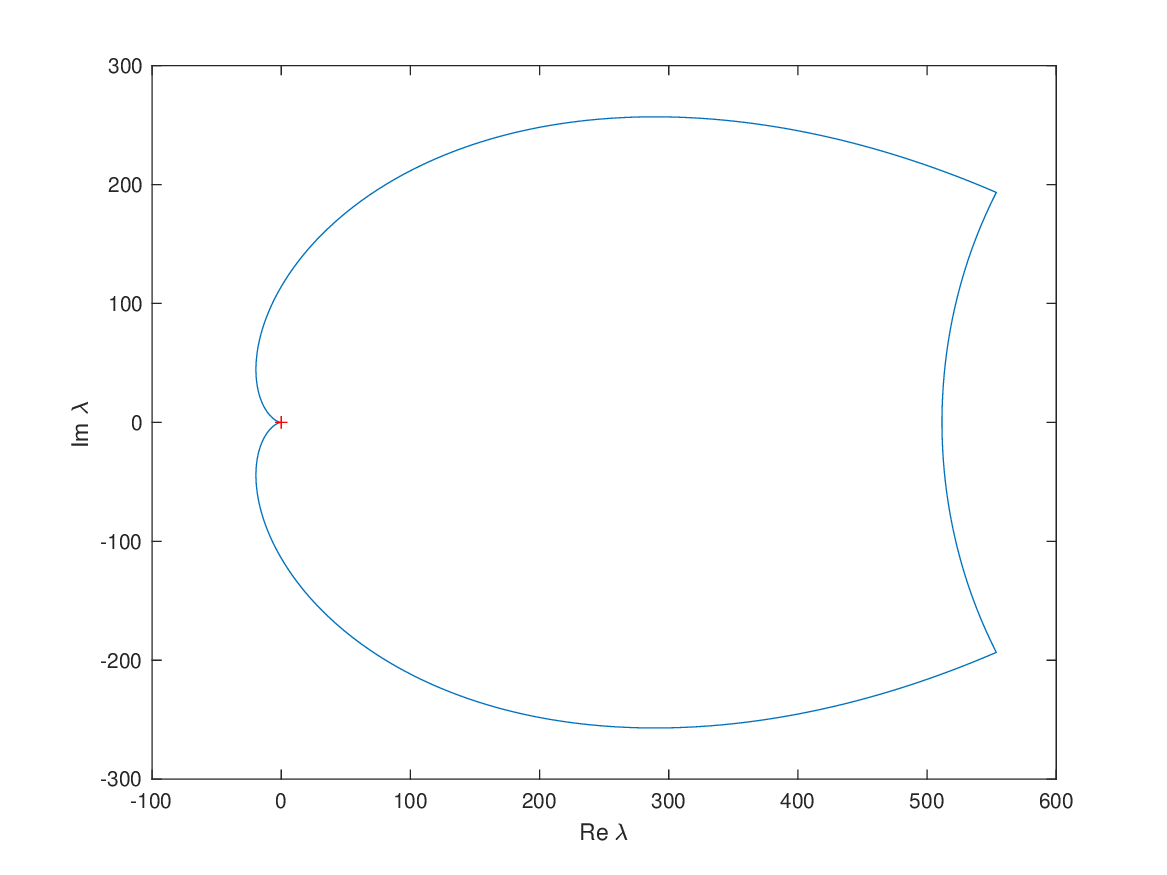}
\end{center}
\caption{The image of a semi-annular region with radii $20$ and $10^{-6}$ through the Evans function $E(\lambda)$ for the QHD system with nonlinear viscosity. The origin is marked in red.}
\label{fig_evans_function_nonlinear_1}
\end{figure}

\begin{figure}
\begin{center}
\includegraphics[scale=0.6]{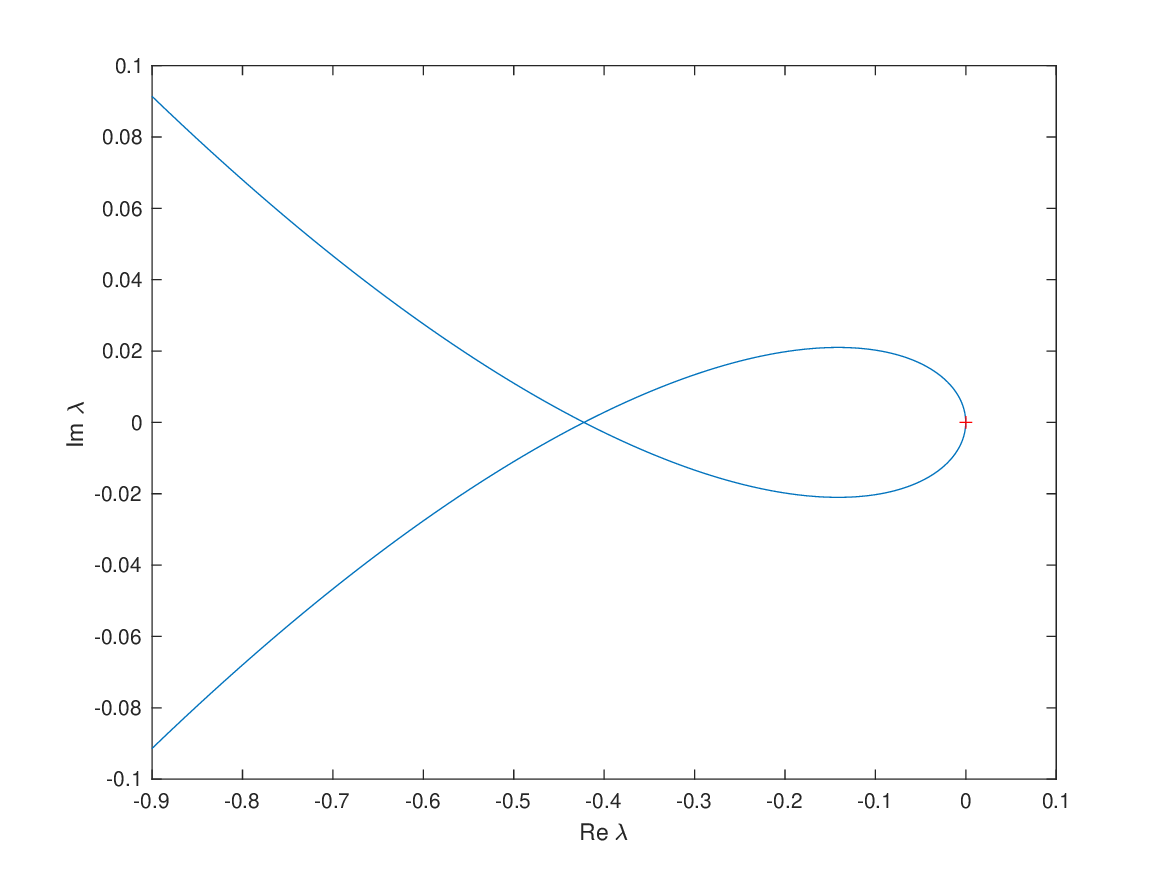}
\end{center}
\caption{Same as Figure \ref{fig_evans_function_nonlinear_1} with enlargement of a region around the origin.}
\label{fig_evans_function_nonlinear_2}
\end{figure}

\begin{figure}
\begin{center}
\includegraphics[scale=0.6]{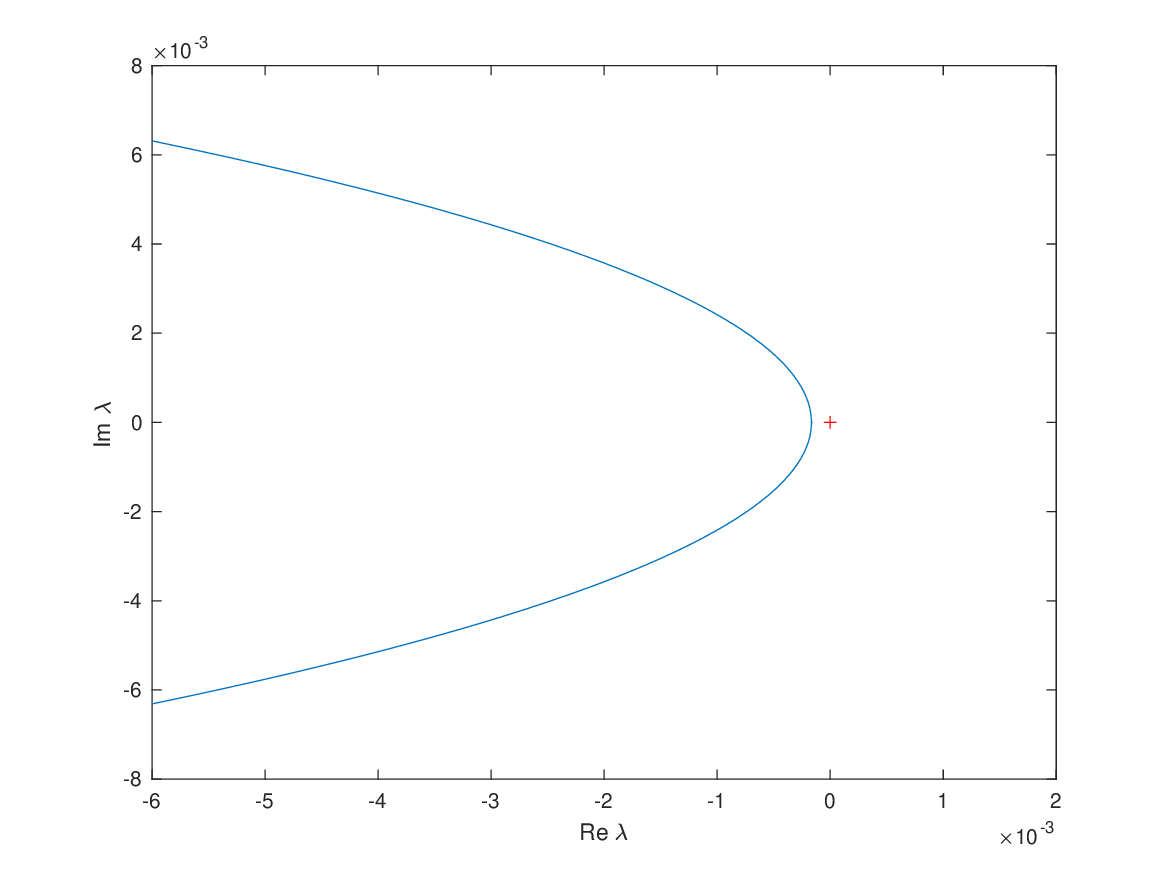}
\end{center}
\caption{Same as Figures \ref{fig_evans_function_nonlinear_1} and \ref{fig_evans_function_nonlinear_2} with further  enlargement of a region around the origin.}
\label{fig_evans_function_nonlinear_3}
\end{figure}

We evaluate the Evans function on the contour surrounding a semi-annular region with radii 20 and $10^{-6}$ with center at the origin and vertical segment on the imaginary axis. We integrate the reduced Kato ODE using $10^6$ points, then we compute $E(\lambda)$ with the solver ode15s in Matlab with relative tolerances $10^{-8}$ and $10^{-10}$, the latter being used if $E(\lambda)$ is evaluated wtih $\lambda$ close to 0. The Evans function $E(\lambda)$ is ploted in Figure \ref{fig_evans_function_nonlinear_1} and in Figures \ref{fig_evans_function_nonlinear_2} and \ref{fig_evans_function_nonlinear_3} with enlargement of a region around the origin. Notice that the Evans function $E(\lambda)$ is small in absolute value for $\lambda \approx 0$ along the contour, hence a higher accuracy is required for its numerical computation than the one used for the QHD system with linear viscosity (cf. Section \ref{sec:numerics_evans_function_linear}). The winding number of the Evans function is (approximately) 1. As before, this is numerical evidence for the presence of a simple real eigenvalue in the interval $(0,20)$.

\begin{figure}
\begin{center}
\includegraphics[scale=0.6]{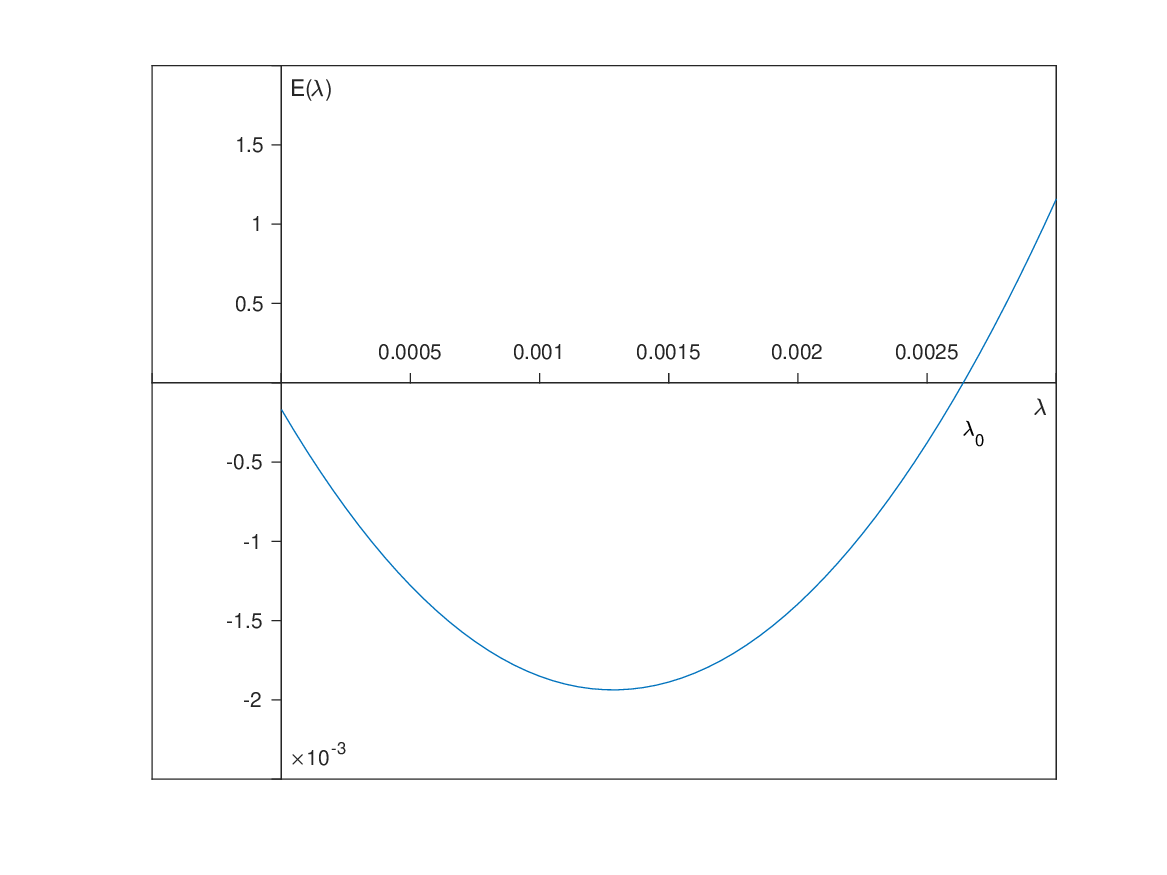}
\end{center}
\caption{The Evans function $E(\lambda)$ computed on a segment of the real line for the QHD system with nonlinear viscosity. It has a zero at $\lambda_0 \approx 0.0026$.}
\label{fig_evans_function_real_nonlinear}
\end{figure}

Now, let us evaluate $E(\lambda)$ on the real line. We initialize the Kato ODE at $\lambda = 20$ and integrate it along the real line until $\lambda = 10^{-6}$ using $4 \cdot 10^5$ points. The result is depicted in Figure \ref{fig_evans_function_real_nonlinear}. The Evans function $E(\lambda)$ has a real root $\lambda_0 \approx 0.0026$. Notice that both the viscosity coefficient and the unstable eigenvalue are smaller in the nonlinear viscosity case, compared to their counterparts in case with the linear viscosity (cf. the parameter set \eqref{parameter_set}), even though increasing the viscosity tends to stabilize the system. Therefore, our numerics suggests that the nonlinear viscosity has stabilizing effect, corroborating the analytical result in \cite{FPZ23} (see Remark 4.7).

\section{Discussion}
\label{sec:discussion}
It was shown numerically in Sections \ref{sec:numerics_evans_function_linear} and \ref{sec:numerical_evans_function_nonlinear} that the standing wave profiles considered for the QHD systems \eqref{qhd_linear_viscosity} and \eqref{qhd_nonlinear_viscosity_cons} with linear and nonlinear viscosity are spectrally unstable. The essential spectrum is stable, however there is a simple real unstable eigenvalue. This numerical result is in contrast with the numerical stability results for traveling wave profiles in \cite{LMZ2020} and \cite{LZ2021} where it was shown that non-monotone traveling wave profiles are spectrally stable. We notice that in both cases there exists a supersonic region, where the velocity along the profile is larger than the speed of sound (see Figures \ref{fig_standing_wave} and \ref{fig_standing_wave2}). On the other hand, one can show numerically that the velocity along the profiles is subsonic everywhere for the cases considered in \cite{LMZ2020} and \cite{LZ2021}. Therefore, we make the following
\begin{conjecture}
\label{Conj:stability_subsonic_or_sonic_profiles}
A standing or traveling wave profile for the QHD systems with linear or nonlinear viscosity \eqref{qhd_linear_viscosity} and \eqref{qhd_nonlinear_viscosity_cons} is spectrally stable if and only if the velocity is subsonic or sonic along the profile.
\end{conjecture}
In other words, the conjecture states that a profile $(R,U)$ is spectrally stable if and only if
\begin{equation*}
|U(x)| \leq c_s(R(x)),\mbox{ }x \in \R.
\end{equation*}
If this inequality holds, then taking limits as $x \rightarrow \pm \infty$ we obtain $|U^{\pm}| \leq c_s(R^{\pm})$, that is the end states are subsonic or sonic, in accordance with the necessary and sufficient condition in Theorem \ref{Thm:essential_spectrum_linear} for stability of the essential spectrum. Let us now compare the numerical results for the QHD systems with linear and nonlinear viscosity. The unstable eigenvalue is larger for the former system even though the viscosity coefficient is smaller for the latter (cf. the parameter sets \eqref{parameter_set} and \eqref{parameter_set_nonlinear_viscosity}). Therefore, our numerical results suggest that the nonlinear viscosity has stabilizing effect.

\section*{Acknowledgement}
This research work was conducted in Gran Sasso Science Institute, University of L'Aquila and Universidad Nacional Aut\'{o}noma de M\'{e}xico.


\begin{thebibliography}{10}
\bibitem{AM} P. Antonelli, P. Marcati, 
On the finite energy weak solutions to a system in Quantum Fluid Dynamics, \textit{Comm. Math. Phys.} 287, 657-686 (2009)

\bibitem{AMZ} P. Antonelli, P. Marcati, H. Zheng, Genuine Hydrodynamic Analysis to the 1-D QHD System: Existence, Dispersion and Stability, \textit{Comm. Math. Phys.} 383, 2113-2161 (2021)

\bibitem{AMZ1} P. Antonelli, P. Marcati, H. Zheng, An intrinsically hydrodynamic approach to multidimensional QHD systems, \textit{Arch. Ration. Mech. Anal.} 247, 24 (2023)

\bibitem{Boh52a} D. Bohm, A suggested interpretation of the quantum theory in terms of ``hidden'' variables. {I}, Phys. Rev. (2), 85, 166-179 (1952)

\bibitem{Boh52b} D. Bohm, A suggested interpretation of the quantum theory in terms of ``hidden'' variables. {II}, Phys. Rev. (2) 85, 180-193 (1952)

\bibitem{BHK87} D. Bohm, B. J. Hiley, and P. N. Kaloyerou, An ontological basis for the quantum theory, Phys. Rep. 144 , no. 6, 321-375 (1987)

\bibitem{Solitons} S. Burger, K. Bongs, S. Dettmer, W. Ertmer, K. Sengstock, A. Sanpera et al., Dark Solitons in Bose-Einstein Condensates,  \textit{Phys. Rev. Lett.} 83, 5198-5201 (1999)

\bibitem{FPZ22} R. Folino, R. G. Plaza, and D. Zhelyazov, Spectral stability of small-amplitude dispersive shocks in quantum hydrodynamics with viscosity, \textit{Commun. Pure Appl. Anal.} 2, no. 12, 4019-4040 (2022).

\bibitem{FPZ23} R. Folino, R. G. Plaza, and D. Zhelyazov, Spectral stability of weak dispersive shock profiles for quantum hydrodynamics with nonlinear viscosity, \textit{J. Differ. Equ.} 359, 330-364 (2023).

\bibitem{Gasser} I. Gasser, Traveling Wave Solutions for a Quantum Hydrodynamic Model, \textit{Applied Mathematics Letters} 14, 279-283 (2001)

\bibitem{Gurevich} A. V. Gurevich and A. P. Meshcherkin. Expanding self-similar discontinuities and shock waves in dispersive hydrodynamics, \textit{Sov. Phys. JETP}, 60(4), 732-740 (1984)

\bibitem{Gurevich1} A. V. Gurevich and L. P. Pitaevskii, Nonstationary structure of a collisionless shock wave, \textit{Sov. Phys. JETP}, 38:291-297 (1974)

\bibitem{Hoefer} M. A. Hoefer, M. J. Ablowitz, I. Coddington, E. A. Cornell, P. Engels, and V. Schweikhard, Dispersive and classical shock waves in Bose-Einstein condensates and gas dynamics, \textit{Phys. Rev. A}, 74, 023623 (2006)

\bibitem{Humpherys} J. Humpherys, On the shock wave spectrum for isentropic gas dynamics with capillarity, \textit{J. Differential Equations}, 246(7):2938-2957 (2009)

\bibitem{Khalatnikov} I. M. Khalatnikov,  \textit{An Introduction to the Theory of Superfluidity}, CRC Press, 2000

\bibitem{Zhelyazov} C. Lattanzio, P. Marcati, D. Zhelyazov, Dispersive shocks in quantum hydrodynamics with viscosity, \textit{Phys.\ D} 402, 132222 (2020)

\bibitem{LMZ2020} C. Lattanzio, P. Marcati, D. Zhelyazov, Numerical investigations of dispersive shocks and spectral analysis for linearized quantum hydrodynamics, \textit{Appl. Math. Comput.} 385, 125450 (2020)

\bibitem{LZ2021} C. Lattanzio, D. Zhelyazov, Spectral analysis of dispersive shocks for quantum hydrodynamics with nonlinear viscosity, \textit{Math. Models Methods Appl. Sci.} 31, no. 9, 1719-1747 (2021)

\bibitem{LZ} C. Lattanzio, D. Zhelyazov, Traveling waves for quantum hydrodynamics with nonlinear viscosity, \textit{J. Math. Anal. Appl.} 493, no. 1, 124503 (2021)

\bibitem{Nov} S. Novikov, S. V. Manakov, L. P. Pitaevskii, and V. E. Zakharov, \textit{Theory of Solitons}, Consultants Bureau, New York, 1984.

\bibitem{PlZ} R. Plaza, D. Zhelyazov, Well-posedness and decay structure of a quantum hydrodynamics system with Bohm potential and linear viscosity, preprint, arXiv:2309.00175

\bibitem{Sagdeev} S. R. Z. Sagdeev,  Kollektivnye protsessy i udarnye volny v razrezhennol plazme (Collective processes and shock waves in a tenuous plasma), in: \textit{Voprosy teorii plazmy (Problems of Plasma Theory)}, Vol. 5, Atomizdat, 1964.

\bibitem{Sandstede} B. Sandstede, \emph{Stability of Travelling Waves}, Handbook of Dynamical Systems II, Elsevier (2002) 983-1055

\bibitem{Zhelyazov1} D. Zhelyazov, Existence of standing and traveling waves in quantum hydrodynamics with viscosity, \textit{Z. Anal. Anwend.} 42, no. 1/2, 65-89 (2023)

\bibitem{Zumbrun} K. Zumbrun, A local greedy algorithm and higher order extensions for global numerical continuation of analytically varying subspaces, \textit{Quart. Appl. Math.} Vol. 68, No. 3, pp. 557-561 (2010)
\end{thebibliography}
\end{document}